\documentclass[aps,twocolumn,showpacs,floatfix,longbibliography]{revtex4-2}
\usepackage{chngcntr}
\usepackage{mathrsfs,braket}
\usepackage{multirow}
\usepackage{times}
\usepackage{amssymb, amsbsy, amsmath, latexsym, dsfont, array, layout, graphicx, mathrsfs, color, ulem, bm}
\usepackage[colorlinks=true,linkcolor=blue,anchorcolor=red,citecolor=blue, urlcolor=blue]{hyperref}

\begin{document}
\title{Time-boundary scattering and topological resonant transmissions}
\author{Haiping Hu}
\email{hhu@iphy.ac.cn}
\affiliation{Beijing National Laboratory for Condensed Matter Physics, Institute of Physics, Chinese Academy of Sciences, Beijing 100190, China}
\affiliation{School of Physical Sciences, University of Chinese Academy of Sciences, Beijing 100049, China}

\begin{abstract}
Time boundaries (TBs), temporal analogues of spatial interfaces, offer a powerful handle to engineer quantum systems. However, unlike the well-developed stationary scattering theory at spatial interfaces, a unified framework for quantum scattering at TBs has been missing. Here we develop a Bloch-wave scattering theory for TBs by introducing a temporal scattering matrix $S$ between incoming and outgoing Bloch channels. We uncover topological resonant transmissions (RTs)—poles of $S$ that yield perfect interband transmission and dynamical freezing of the quantum state. We establish a bulk–time–boundary correspondence for all integer Altland-Zirnbauer classes: the number of RTs equals the jump of the bulk topological invariant across the TB. In one dimension this gives a time-domain Levinson’s theorem. A topological analysis further reveals a striking dimensional dependence. In even dimensions RTs are robust to temporal modulations and disorder, whereas in odd dimensions they can be destroyed by dynamical symmetry breaking. Our work places temporal and spatial scattering on the same footing and opens new avenues for engineering and probing quantum dynamics.
\end{abstract}
\maketitle


Classical waves crossing a spatial interface undergo reflection and refraction, as dictated by Fermat’s principle. A time boundary (TB) is the temporal analogue of a spatial interface, realized for example in optics as an abrupt change of the refractive index in time. The space–time duality of Maxwell’s equations implies that waves encountering TBs experience time reflection and refraction \cite{duality1,duality2,duality3,duality4,duality5,duality6}. Unlike spatial boundaries which conserve energy but alter momentum, a TB conserves momentum while changing energy due to broken time-translation symmetry. Classical TB effects have been observed in microwave and optical systems \cite{mwTB1,mwTB2,mwTB3,opticTB1,opticTB2}, acoustics \cite{TB5}, and water waves \cite{TB6}, with applications in broadband frequency translation \cite{opticTB1,mwTB3}, nonreciprocal devices \cite{TB7}, and photonic time crystals \cite{TB8,TB9,TB10,TB11,TB12}.

In the quantum realm, wave propagation obeys the Schrödinger equation, $i\partial_t|\Psi(t)\rangle=H|\Psi(t)\rangle$. Reflection and transmission at a spatial interface are governed by stationary scattering theory. The simplest TB is a sudden quench of the Hamiltonian in time, which generates a wealth of nonequilibrium phenomena and is widely used as a tool for quantum control. Extensive work on quench dynamics has revealed rich spatiotemporal topological patterns \cite{qtopo7,quench_vortice,qtopo13,qtopo14,qtopo15,qtopo1,qtopo2,qtopo3,qtopo5,qtopo6,qtopo8a,nigel_symmetry,qtopo9,qtopo10,qtopo11,qtopo12,qtopo8c,qtopo17,qtopo18,qtopo19,qtopo8b,qtopo20,nigel_class} for topological band systems \cite{topo_rev1,topo_rev2}. 
In most of these studies the focus is on the the evolution of quantum states and dynamical observables. The quench is treated primarily as a ``black-box’’ drive for quantum dynamics, rather than as a TB with its own scattering mechanism. Very recently, quench-induced time ``reflection’’/refraction was observed in a one-dimensional (1D) Su–Schrieffer–Heeger lattice of cold atoms \cite{yanbo}. Yet quantum scattering at TBs remains far less understood \textit{viz-à-vis} its well-established spatial counterpart. Unlike classical waves, quantum spatial and temporal scattering breaks space–time duality, since the Schrödinger equation is second order in space but only first order in time. Moreover, core scattering notions at TBs, including a well-defined scattering matrix, its poles, and their relation to spatial scattering and topology, have remained elusive.

From a practical standpoint, most previous studies of TBs in classical and quantum systems have focused on sharp, abrupt parameter changes, which are experimentally demanding and restrictive. In reality, TBs can be much more flexible, generic, and experimentally friendly. For instance, they can have finite duration, or follow smooth parameter ramps \cite{kibble,zurek} or periodic modulations as in photonic time crystals \cite{TB8,TB9,TB10,TB11,TB12}. Such versatile forms of temporal engineering are now routine in platforms like cold atoms \cite{drive1,drive2,topo_rev3,topo_rev4} and photonic materials \cite{LuNatPhoton2014,OzawaRMP2019} and serve as a powerful tool for quantum control. Taken together, these developments call for a comprehensive theory for temporal scattering at generic TBs that unveils its fundamental dynamical principles and intrinsic connections to the topology of quantum systems.

In this work, we develop a unified scattering framework for Bloch waves at generic TBs. By treating the TB as a scattering object, we formulate a temporal scattering matrix $S$ that relates incoming and outgoing Bloch waves across the boundary. We identify topological resonant transmissions (RTs) as poles of the $S$-matrix that give rise to perfect inter-channel transmission, diminish birefringence, and dynamically freeze the quantum state. For all integer Altland–Zirnbauer (AZ) classes \cite{class1,class2,class3}, we establish a bulk–time–boundary correspondence in which the number of RT modes equals the change of the bulk topological invariant across the TB; in 1D this correspondence yields a time-domain Levinson’s theorem, a fundamental relation between scattering phase shifts and ``bound states’’. A topological analysis further shows that RT robustness hinges on spatial dimensions. In even dimensions RTs are stable against temporal modulations and disorder, whereas in odd dimensions they may break down through dynamical symmetry breaking within the TB. Our results unify temporal and spatial scattering in quantum systems and offer fresh insights into dynamical topological phenomena.\\

\noindent{\large{\bf Results}}

\noindent{\bf Scattering matrix at a time boundary.}~We consider a $d$D lattice system with time-varying parameters. A generic setting of TB is specified by a piecewise time-dependent Hamiltonian
\begin{equation}
H(t)=
\begin{cases}
H^L, & t\leq0,\\[2pt]
H^{\rm TB}(t), & 0< t< \tau_B,\\[2pt]
H^R, & t\geq\tau_B,
\end{cases}
\end{equation}
where $H^L$ and $H^R$ are static Hamiltonians before and after the TB, and $\tau_B$ is the duration of the TB. The special case $\tau_{B}=0$ corresponds to a sudden quantum quench. Inside the TB, $H^{\rm TB}(t)$ is otherwise arbitrary. The lattice system has $2N$ bands. The Bloch eigenstates and energy bands of $H^{L,R}$ (ordered from low to high) are denoted by $|\psi_j^{L,R}(\bm k)\rangle$ and $E^{L,R}_j(\bm k)$ ($j=1,2,\dots,2N$), with $\bm k=(k_1,k_2,\dots,k_d)$ the crystal momentum. We set the Fermi level to zero, which can always be done for gapped topological phases. Bloch bands below (above) it, $E^{L,R}_j(\bm k)$ with $j=1,2,\dots,N$ ($j=N+1,\dots,2N$), are valence (conduction) bands with negative (positive) energies. Time evolution across the TB obeys the Schrödinger equation. The general solutions before and after the TB can be written as
\begin{align}
&|\Psi_{\textrm{before}}\rangle = \sum_{j=1}^{N} a_{j} e^{i |E^L_j| t} |\psi_j^L\rangle+\sum_{j=N+1}^{2N} a_{j} e^{-i E^L_j t} |\psi_j^L\rangle, \notag \\
&|\Psi_{\textrm{after}}\rangle = \sum_{j=1}^{N} b_{j} e^{i |E^R_j| t} |\psi_j^R\rangle+\sum_{j=N+1}^{2N} b_{j} e^{-i E^R_j t} |\psi_j^R\rangle. \notag
\end{align}

\begin{figure}[!t]
\centering
\includegraphics[width=3.33 in]{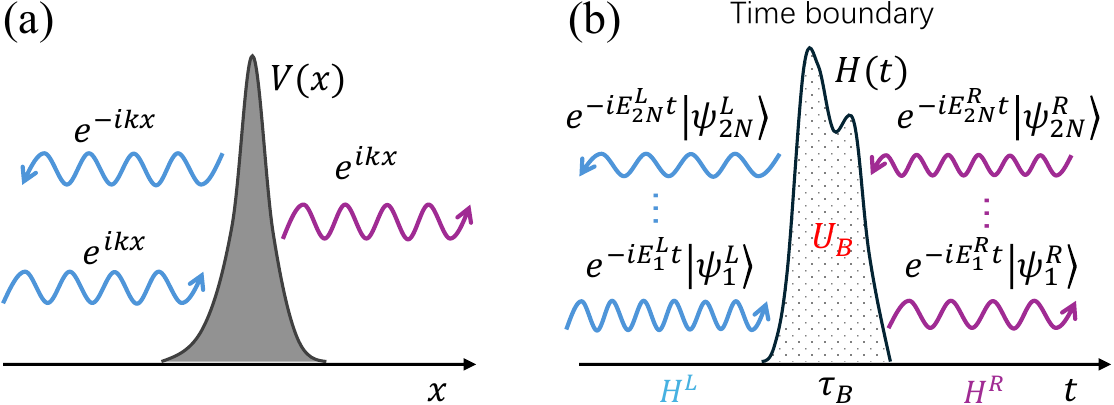}
\caption{Spatial versus temporal scattering. (a) In real space, a plane wave of fixed energy impinges on a potential $V(x)$, leading to reflected and transmitted waves and possibly bound states. The process is governed by stationary scattering theory. (b) In time, the Hamiltonian changes from $H^L$ to $H^R$ across a time boundary (TB). Bloch waves of definite crystal momentum act as scattering channels, while energy is not conserved. A TB of duration $\tau_B$ generates a unitary evolution $U_B$ and the scattering obeys the Schrödinger equation. Arrows pointing towards (away from) the TB indicate incoming (outgoing) waves, and colors distinguish states before and after the boundary.}\label{fig1}
\end{figure}
Instead of focusing on the full time evolution and nonequilibrium quantum dynamics, we treat the TB as a genuine scattering interface between two static bulk phases with well-defined asymptotic Bloch bands at $t\leq0$ and $t\geq\tau_B$. This conceptual shift from time evolution to time scattering is crucial for the results that follow. Figure \ref{fig1} draws an analogy between 1D spatial scattering and temporal scattering. In the former, a plane wave $e^{ikx}$ of fixed energy scatters off a potential $V(x)$, producing reflected and transmitted waves. In temporal scattering, the crystal momentum $\bm k$ is conserved across the TB but energy is not. Excitations are produced as the system passes through the TB. The Bloch eigenstates act as $2N$ distinct scattering channels. Formally, time $t$ plays the role of spatial coordinate $x$, and the plane‑wave phase factor is replaced by a dynamical phase:
\begin{eqnarray}\label{analogy}
x \leftrightarrow t; \quad e^{ikx} \leftrightarrow e^{-i E^{L,R}_j t}.
\end{eqnarray}
During scattering, the initial wave $|\Psi_{\textrm{before}}\rangle$, expressed in the eigenbasis of $H^L$, enters the TB. Upon exit, it becomes $|\Psi_{\textrm{after}}\rangle$. The temporal transfer matrix $M$ is defined by
\begin{eqnarray}
(b_1, b_2, \dots, b_{2N})^T = M~(a_1, a_2, \dots, a_{2N})^T. 
\end{eqnarray}
As the scattering process follows the Schrödinger equation, the TB induces a unitary evolution $U_{B}=\mathbb{T}e^{-i\int_0^{\tau_{B}}H^{TB}(t)dt}$ ($\mathbb{T}$ means time ordering). The elements of $M$ are
\begin{eqnarray}\label{utransform}
M_{mn} = \langle \psi_m^R | U_{B} | \psi_n^L \rangle.
\end{eqnarray}
$U_{B} =1$ for a sudden quench. The $M$-matrix, determined by $H^{L}$, $H^{R}$ and the specific profile of $H^{TB}(t)$, encodes how an initial Bloch wave scatters at the boundary. Its unitarity, $M^\dagger M = M M^\dagger = 1$, ensures probability conservation of the scattering process.

In scattering theory, the $S$-matrix is the central object that relates incoming waves on either side of a boundary to outgoing waves (including both reflected and transmitted components). To obtain it, we follow the analogy in Eq.~(\ref{analogy}). For spatial scattering, an incoming plane wave $e^{ikx}$ is partially reflected into $e^{-ikx}$ upon encountering a potential. In temporal scattering, Bloch waves with negative (positive) energy before the TB are treated as incident (reflected) waves, with the roles reversed after the TB, as shown in Fig. \ref{fig1}(b). Note that in this scattering formulation, the ``reflected'' waves differ from the ``reflected'' components discussed in Refs.~\cite{yanbo,qtopo19,qtopo20}, which refer to waves that reverse their group velocity after crossing the TB (i.e., spatially propagate in the opposite direction). The $S$-matrix then reads
\begin{eqnarray}
(b_{1\rightarrow N};a_{N+1\rightarrow 2N})^T=S~(a_{1\rightarrow N};b_{N+1\rightarrow 2N})^T.
\end{eqnarray}
A straightforward calculation yields
\begin{equation}
S =
\begin{pmatrix}
M_{--} - M_{-+} M_{++}^{-1} M_{+-} & M_{-+} M_{++}^{-1} \\
-\,M_{++}^{-1} M_{+-} & M_{++}^{-1}
\end{pmatrix},
\label{Smatrix}
\end{equation}
where $M_{--}$, $M_{-+}$, $M_{+-}$, and $M_{++}$ denote the upper-left, upper-right, lower-left, and lower-right blocks of the $M$-matrix, respectively. The matrix elements of $S$ encode the reflection and transmission amplitudes between different channels, and the eigenphases of $S$ represent the associated scattering phase shifts.\\

\noindent{\bf Resonant transmission.}~The $S$-matrix allows us to analyze the scattering process at the TB in a precise manner. In the multichannel scattering, an incident Bloch wave with crystal momentum $\bm k$ generally splits, after crossing the TB, into $2N$ outgoing branches of $H^R$ ($N$ in the conduction sector and $N$ in the valence sector). In two-band systems, this manifests as birefringence. We are particularly interested in momenta at which the TB acts as a perfect ``frequency converter” between Bloch bands, converting an incoming state in one band entirely into an outgoing state in another. Within our scattering framework, these special points appear as poles of the time-domain $S$-matrix, which we term topological resonant transmissions (RTs). In scattering theory, poles of the $S$-matrix are particularly important: in stationary scattering, poles on the real-energy axis correspond to bound states. This naturally raises the question of what physical information such poles encode in the TB scattering, what nontrivial effects they generate, and how they are related to the topology. From the explicit form of the $S$-matrix in Eq.~(\ref{Smatrix}), the poles occur when $M_{++}$ is non-invertible, i.e., $\det M_{++}(\bm k)=0$. Near a pole $\bm k_0$, the norm diverges as $\|M^{-1}_{++}(\bm k)\|\sim|\bm k-\bm k_0|^{-p}$, where $p$ is the order of the pole. [See Appendix \textbf{S1} for examples of higher-order poles.] 
\begin{figure}[!t]
\centering
\includegraphics[width=3.33 in]{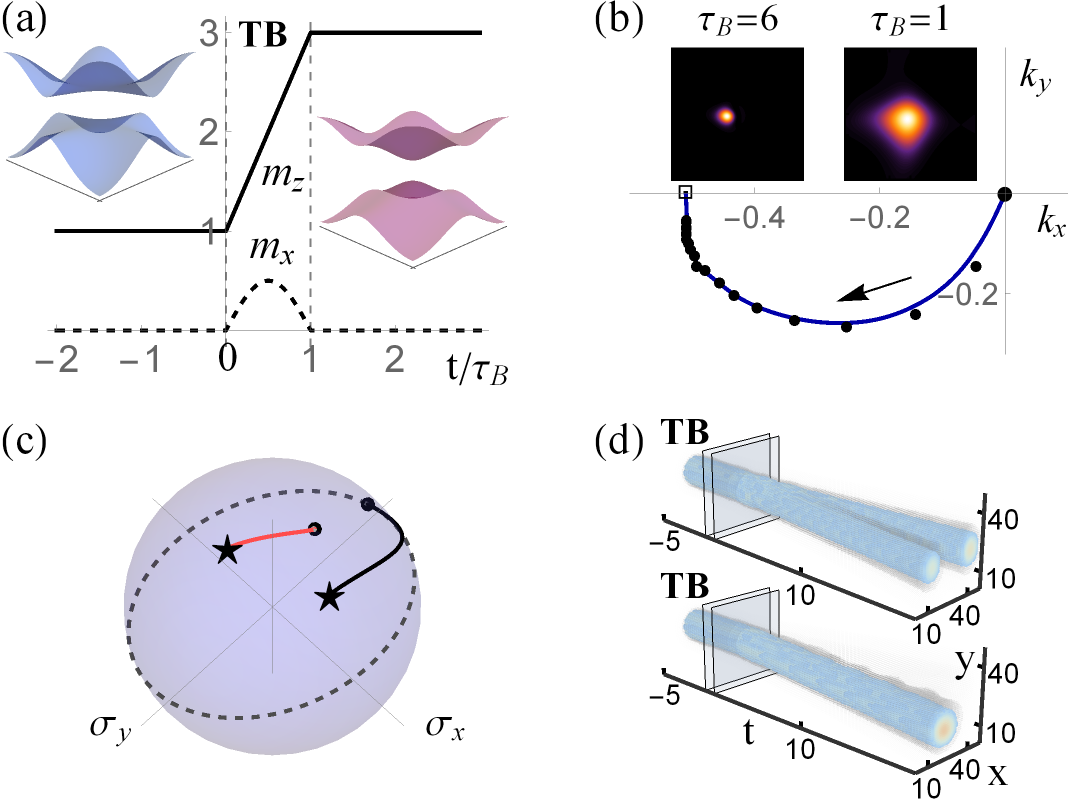}
\caption{Resonant transmission (RT) across a TB for the 2D model in Eq. (\ref{2dqwz}). (a) Schematic of the TB with time-varying mass terms $m_x(t)$ and $m_z(t)$. Band structures on either side are shown in blue and purple. (b) Trajectory of the poles of the $S$-matrix in the Brillouin zone (BZ) as the TB duration $\tau_B$ increases (indicated by the arrow). Contour plots show the inter-channel transmission $T_{-+}$ for two different $\tau_B$ values. The brightest points mark the RT. (c) Time evolution on the Bloch sphere for the RT mode (red) and an off-resonant mode $\bm k=(0.5,0)$ (black). The system starts in the ground state of $H^L$ (black stars). The evolution inside and after the TB is shown by the solid and dashed lines, respectively, with the black dots marking the points when the state exits the TB. (d) Scattering of Gaussian wave packets. (Top) $\bm{k} = (0.5, 0)$; (Bottom) RT mode, $\bm{k} = (-0.144, -0.24)$. Parameters: packet center at $\bm r_0=(25,25)$, width $\sigma=6$, and system size $L_x = L_y = 50$. $\tau_B=1$ for (c)(d).}\label{fig2}
\end{figure}

We illustrate the TB scattering and the emergence of RTs using a 2D two-band lattice model. The time-dependent Hamiltonian is
\begin{equation}\label{2dqwz}
H(\bm k,t)=\bm h(\bm k)\cdot\bm\sigma + m_x(t)\sigma_x + m_z(t)\sigma_z,
\end{equation}
where $\bm h(\bm k) = \bigl(\sin k_x,\sin k_y,-\cos k_x-\cos k_y\bigr)$ and $\bm\sigma$ denotes the Pauli matrices. The TB is implemented over the interval $t\in(0,\tau_B)$, with time-varying mass terms $m_x(t)=\tfrac{1}{2}\sin\frac{t}{\tau_B}$ and $m_z(t)=\Bigl(1-\frac{t}{\tau_B}\Bigr)m^L+\frac{t}{\tau_B}m^R$, as shown in Fig.~\ref{fig2}(a). On either side of the TB, the Hamiltonian reduces to the paradigmatic Qi–Wu–Zhang model of a Chern insulator \cite{qwzmodel}, with mass $m_z=m^L$ and $m_z=m^R$, respectively. The band topology is characterized by the Chern number, $C = \mathrm{sgn}(m_z)$ for $0<|m_z|<2$ and $C=0$ otherwise. In what follows, we fix $m^L=1$ and $m^R=3$, so that the band topology changes from $C=1$ to $C=0$ across the TB.

For each fixed TB duration $\tau_B$, there is a single pole in the 2D BZ. Its trajectory as a function of $\tau_B$ is shown in Fig.~\ref{fig2}(b). At $\tau_B=0$ (a quenched TB), the pole sits at the zone center $\bm k=(0,0)$; as $\tau_B$ increases, it gradually moves in the BZ. No pole appears if $H^L$ and $H^R$ lie in the same topological phase. We next examine the momentum-dependent transmission across the TB. Starting from the ground state of $H^L$, $|\psi^L_-(\bm k)\rangle$ at $t=0$, the TB scatters it into both eigenchannels of $H^R$. The inter-channel transmission is characterized by
\begin{equation}
T_{-+}(\bm k) = \bigl|\langle\psi^L_-(\bm k)\,|\,\psi^R_+(\bm k)\rangle\bigr|^2 .
\end{equation}
Figure~\ref{fig2}(b) shows the distribution of $ T_{-+}(\bm k)$ in the BZ for two representative TB durations. One clearly sees that perfect transfer occurs at RT mode. As $\tau_B$ increases, the resonant peak becomes sharper but never vanishes.

A key physical consequence of RT is the dynamical freezing of the quantum state in time evolution. Consider the time-evolved state $|\psi(t)\rangle = \mathbb{T}e^{-i\int_0^{t}H(t)dt}|\psi_-^L(\bm k)\rangle$. For a two-level system, this state can be represented by a single point
$\langle\psi(t)|\bm\sigma|\psi(t)\rangle$ on the Bloch sphere. As shown in Fig.~\ref{fig2}(c), for the RT mode, the state moves on the Bloch sphere when it is inside the TB. Upon exit, it no longer evolves and instead becomes frozen at a fixed point on the sphere. In contrast, for a typical off-resonant mode, the evolution persists as Larmor precession after exiting the TB. These differences are also reflected in the wave-packet dynamics, as shown in Fig.~\ref{fig2}(d). We prepare a Gaussian wave packet with central momentum $\bm k$, as can be realized in cold-atom experiments~\cite{yanbo}. The initial wave function is $|\psi(\bm r,t=0)\rangle = \eta(\bm r)\otimes|\psi^L_-(\bm k)\rangle$, where
\begin{equation}
\eta(\bm r)\propto
e^{-\frac{(\bm r-\bm r_0)^2}{2\sigma^2}}e^{i\bm k\cdot(\bm r-\bm r_0)}
\end{equation}
describes an envelope centered at $\bm r_0=(x_0,y_0)$ with width $\sigma$. For an off-resonant $\bm k$, the wave packet exhibits birefringence, splitting into two branches with opposite group velocities. By contrast, for the RT momentum, the packet passes through the TB transparently without birefringence.\\

\noindent{\bf Bulk-time-boundary-correspondence.} The RT can appear in gapped topological phases in any spatial dimension. [See Appendix \textbf{S2} for the case of 3D chiral topological insulators.] We focus on $d$D integer AZ classes, whose Hamiltonian admits an elementary representation via Clifford algebra:
\begin{equation}
H(\bm{k}) = \sum_{a=0}^{d} h_a(\bm{k})\, \Gamma^a.
\end{equation}
The AZ class is specified by three fundamental symmetries: time-reversal $\mathcal{T}$, particle-hole $\mathcal{P}$, and chiral $\mathcal{C}$. The $\Gamma$ matrices reduce to Pauli matrices in the previous example. Let $\bm h=(h_0, h_1,\cdots, h_d)$, then $\vec{\bm h}=\frac{\bm h}{|\bm h|}$ defines a map from the BZ, a $d$D torus $T^d$, to the $d$-sphere $S^d$. The topological invariant $n$ is the degree of this map, given by the winding or Chern number for odd or even $d$.

\begin{figure}[!t]
\centering
\includegraphics[width=3.33 in]{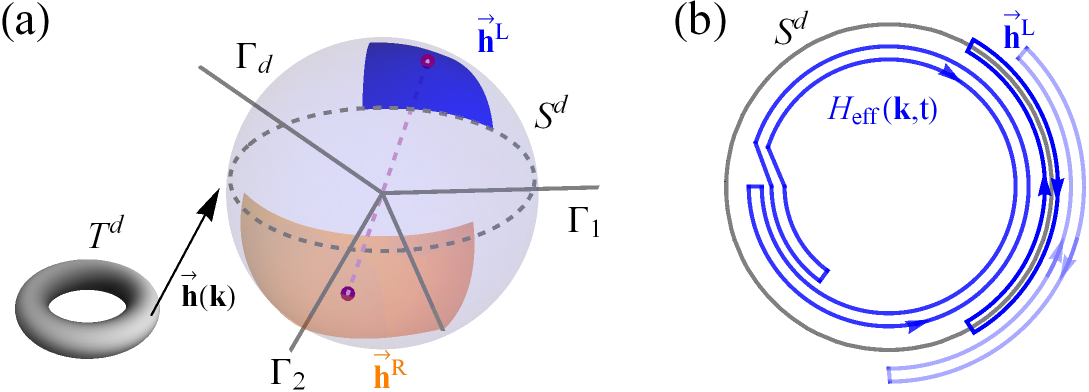}
\caption{Topological view of RT in the elementary Clifford representation. (a) For integer AZ classes in $d$D, the topological invariant is the mapping degree from the Brillouin zone ($T^d$) to the $d$ sphere ($S^d$) spanned by the normalized Hamiltonian vector $\vec{\bm h}$. A RT arises at a quenched TB when $\vec{\bm h}^L$ and $\vec{\bm h}^R$ are antiparallel. (b) Stability of RTs in even dimensions. Temporal modulations inside the TB can stretch the mapping encoded by the effective Hamiltonian $H_{\mathrm{eff}}(\bm k,t)$, but cannot alter the global topology without gap closings.}\label{fig3}
\end{figure}

Remarkably, the number of RT modes is not a microscopic detail but a topological quantity. For the case of quenched TB connecting two phases with topological invariants $n_L$ and $n_R$, one can show that the number of RT modes in the dD BZ, denoted by $N_{RT}$, is given by the difference of topological invariants across the TB:
\begin{eqnarray} \label{rtnumber}
N_{RT} \equiv \sum_{\bm k_0 \in RT} p_{\bm k_0} = |n_L - n_R|.
\end{eqnarray}
Here $p_{\bm k_0}$ is the pole order of the RT at $\bm k_0$. $N_{RT}$ counts all RT modes weighted by their orders. A geometric view of the RT is illustrated in Fig.~\ref{fig3}(a). Consider the interpolation between the two Hamiltonians on either side of the TB: $\bm h_{\lambda}(\bm k) = (1-\lambda)\bm h^L(\bm k) + \lambda\bm h^R(\bm k)$, $\lambda \in [0,1]$. If $H^L$ and $H^R$ differ topologically, the interpolation undergoes $|n_L - n_R|$ gap closings at the Fermi level as $\lambda$ varies from $0$ to $1$. At each closing, $\bm h_\lambda(\bm k) = 0$. This means that the two Hamiltonian vectors $\vec{\bm h}^L$ and $\vec{\bm h}^R$ are antiparallel to each other on the sphere $S^d$. The corresponding $\bm k$  then represents an RT mode where the valence and conduction bands of $H^L$ and $H^R$ exchange. A proof for generic multiband systems, without relying on the elementary Clifford representation, is provided in Methods.

We have seen that the RTs lead to dynamical freezing in the time evolution and result in vanishing birefringence. What is the counterpart of these RTs in the spatial scattering? In stational scattering theory, Levinson’s theorem \cite{Levinson,levinson_book} is a fundamental and elegant relation between the scattering phase shift and the number of bound states $N_{bound}$:
\begin{eqnarray}
N_{bound}=\frac{1}{\pi}[\delta(0)-\delta(\infty)]=\frac{1}{\pi}\left|\int_{0}^{\infty}\frac{\partial \delta(E)}{\partial E}dE\right|,
\end{eqnarray}
where $\delta(E)$ is the scattering phase shift at energy $E$. Physically, the theorem implies that the bound state leaves a fingerprint in the continuum scattering states. For the TB, we identify a dual counting principle for RTs. From the $S$-matrix in Eq.~(\ref{Smatrix}), we have [See Methods for the proof]
\begin{eqnarray}\label{winding1}
N_{RT}=\frac{1}{2\pi}\left|\int_{-\pi}^{\pi}\frac{\partial\delta_{rel}(k)}{\partial k}dk\right|,
\end{eqnarray}
where $\delta_{rel}(k) = \arg[\det M]$ is the relative phase shift between valence and conduction channels. We refer to this as time-domain Levinson’s theorem, given the duality $E \leftrightarrow k$. It states that the winding of relative phase shift yields the number of RTs. Thus, the RT mode is analogous to bound states at spatial interfaces, while off-resonant modes are akin to scattering states. Eq.~(\ref{rtnumber}) can be viewed as a bulk–time–boundary correspondence. The number of RT modes is determined by the jump of the bulk topological invariant across the TB, in direct analogy with domain-wall states at spatial interfaces between topologically distinct phases, where the number is given by the difference of bulk invariants on either side. Despite the duality, we note a key difference: the spatial Levinson’s theorem counts bound states isolated from the continuum spectra, whereas for the TB, the RT modes are dynamically frozen and embeded in the continuum BZ. 

Next, we consider the scenario of generic TB settings and discuss the robustness of RTs against temporal modulations inside the TB. A striking outcome of our analysis is that RT robustness crucially depends on spatial dimensionality. For each AZ class, we assume that the instantaneous Hamiltonian $H^{TB}(t)$ inside the TB preserves the relevant symmetries. By robustness, we mean that the number of RTs remains fixed by the topological difference between $H^L$ and $H^R$, even though the corresponding momentum modes hosting RTs may shift. In such generic settings, an additional unitary transformation acts on $H^L$ [see Eq.~(\ref{utransform})]. The time-evolved states $|\psi^L_n(t)\rangle$ within the TB can be regarded as eigenstates of an effective Hamiltonian  
\begin{eqnarray}\label{heff}
H_{\mathrm{eff}}(\bm{k}, t) = U(0,t) H^L(\bm{k}) U^{\dagger}(0,t), \quad t \in (0,\tau_{B}),
\end{eqnarray}
with $U(0,t) = \mathbb{T} e^{-i \int_0^{t} H(t') dt'}$. According to the bulk-time-boundary correspondence in Eq.~(\ref{rtnumber}), the RT is determined by the topological difference between $H^R$ and $H_{\mathrm{eff}}(\bm k,\tau_{B})$. As the spectrum of $H_{\mathrm{eff}}(\bm k, t)$ is fixed (and gapped) inside the TB, its topology is unchanged as long as the relevant symmetries are preserved.

For the three symmetries of all AZ classes, we have: $\mathcal{T} H_{\mathrm{eff}}(\bm{k},t)\mathcal{T}^{-1}\neq H_{\mathrm{eff}}(-\bm{k},t)$; $\mathcal{P} H_{\mathrm{eff}}(\bm{k},t) \mathcal{P}^{-1}=-H_{\mathrm{eff}}(-\bm{k},t)$; $\mathcal{C} H_{\mathrm{eff}}(\bm{k},t) \mathcal{C}^{-1} \neq -H_{\mathrm{eff}}(\bm{k},t)$. Namely, the $\mathcal{T}$ and $\mathcal{C}$ symmetries can be broken by the temporal modulation within the TB, while the $\mathcal{P}$ symmetry can not. It follows that the RTs are robust in class A (no symmetry, $d=2,4,6,8$), D ($\mathcal{P}^2=1$, $d=2,6$) and C ($\mathcal{P}^2=-1$, $d=2,6$). This topological robustness is intuitively visualized in Fig.~\ref{fig3}(b). Although the transformation in Eq.~(\ref{heff}) deforms the mapping $T^n \to S^n$ of $\vec{\bm h}^L$, the topological invariant $n_L$ remains intact without any gap closing, due to the preservation of symmetry for these symmetry classes. In contrast, for classes AIII, BDI, DIII, CII, and CI which possess $\mathcal{C}$ symmetry, the temporal modulation inside the TB can alter the topology of $H_{eff}(\bm k,t)$, rendering the RT unstable. Such change in topology resembles a rubber band adjusting its winding along the equator as it stretches over a sphere. Note that these five classes support integer invariants only in odd $d$, consistent with the fact that the winding numbers depend on the transformation $U(0,t)$. For class AI and AII, the breaking of $\mathcal{T}$ symmetry reduces them to class A. Their Chern number is unaffected by the transformation $U(0,t)$ without gap closings, and the RT is stable. To sum up, the RT is robust for integer AZ classes (A, D, C, AI, AII) in even $d$. In contrast, for odd $d$ (AIII, BDI, DIII, CII, CI), RT stability requires additional constraints evading dynamical breaking of symmetries. \\

\noindent{\bf Robustness against disorder.}~The emergence of RT as a topological effect is tied to the topological change across the TB and should be robust against disorder. Although RT is formulated in terms of Bloch states in momentum space, its physical effects persist even in the presence of disorder, which breaks lattice translation symmetry. To study the interplay between disorder and TB effects, we introduce Anderson-type disorder into the Hamiltonian after the TB, with the disorder term $H_{\text{dis}} = \sum_j w_j c_j^\dagger c_j$, where $w_j \in [-W/2,W/2]$ and $W$ is the disorder strength. We note that the conclusion holds for other disorder models or disorder applied over other temporal stages. The stability of RT is independent of the microscopic details of the disorder, as long as the relevant symmetries are respected.

\begin{figure}[!t]
\centering
\includegraphics[width=3.33 in]{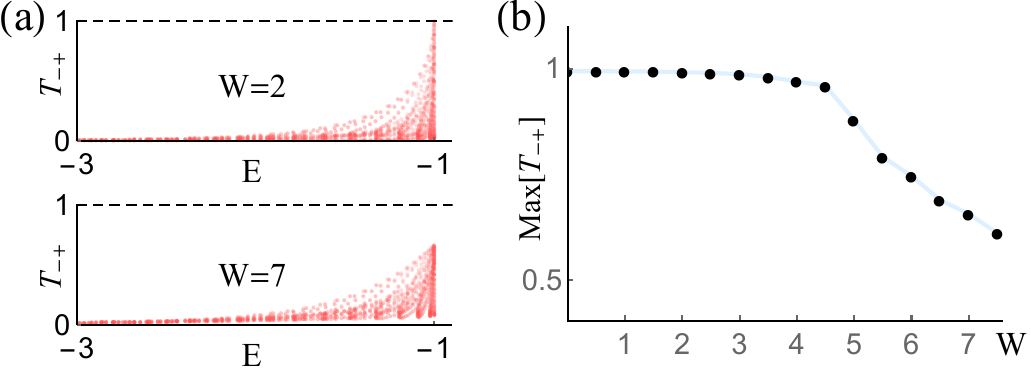}
\caption{Stability of RT in the presence of disorder for model (\ref{2dqwz}). (a) Inter-channel transmission probability $T_{-+}$ versus eigenenergy of $H^L$ for two different disorder strengths, (top) $W=2$ and (bottom) $W=7$. The horizontal dashed line serves as a guide. (b) Maximum inter-channel transmission probability (among all valence-band eigenstates of $H^L$) to the conduction bands of $H^R$ versus the disorder strength $W$.}\label{fig4}
\end{figure}
For disordered systems, we consider eigenstates in the valence (negative-energy) bands of $H^L$, labeled by their eigenenergy $E=E_j^L<0$, and compute its transmission probability $T_{-+}(E)$ to the conduction (positive-energy) bands of $H^R$ after crossing the TB. As shown in Fig.~\ref{fig4}(a), for weak disorder $W=2$, the transmission probability $T_{-+}(E)$ exhibits noticeable fluctuations as a function of the eigenenergy, yet a prominent transmission peak with $T_{-+}(E)\approx 1$ appears at the band edge. This suggests that there is always a certain eigenstate of $H^L$ that, after scattering at the TB, fully transfers to the conduction band of $H^R$. For strong disorder $W=7$, this peak disappears, indicating that sufficiently strong disorder destroys the RT. In Fig.~\ref{fig4}(b), we plot the maximum transmission among all negative-energy eigenstates of $H^L$ as a function of disorder strength $W$. It is evident that perfect transmission survives up to a considerable disorder strength ($W\approx4.5$) and then collapses in the strong disorder regime. These numerical results highlight the topological robustness of RT against disorder.\\

\begin{figure}[!t]
\centering
\includegraphics[width=3.33 in]{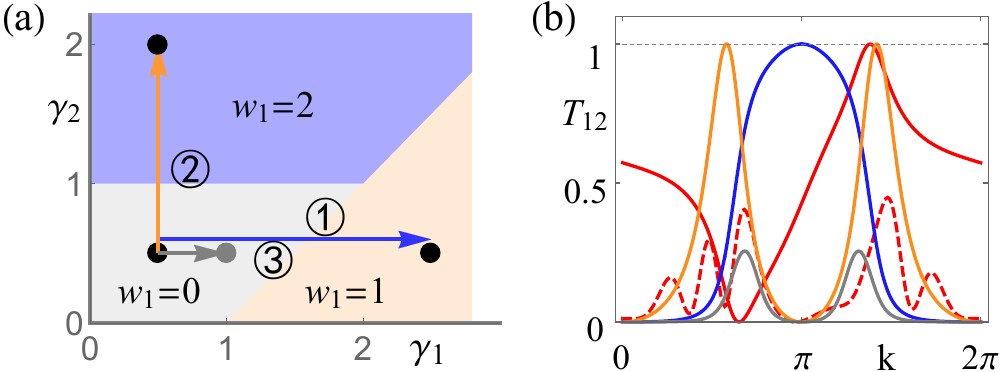}
\caption{RT in the extended SSH model (\ref{1dssh}) and dynamical symmetry breaking. (a) Phase diagram with topologically distinct phases (colored regions) labeled by the winding number $w_1$. Arrows indicate the three quenched TBs. (b) Inter-channel transmission probability $T_{-+}$ versus momentum $k$ for different TB settings. The blue, orange, and gray lines correspond to the quenched protocols \textcircled{1} \textcircled{2} and \textcircled{3} in (a). The red and dashed red lines are for temporal modulation $H^{TB}(t)=\sigma_z$ and $H^{TB}(t)=(1 -t/\tau_B)H^L + t/\tau_B H^R + \sigma_z$, respectively.}\label{fig5}
\end{figure}

\noindent{\bf Dynamical symmetry breaking.}~In odd dimensions, the scattering process at the TB can be accompanied by dynamical symmetry breaking, which disrupts the RT. We demonstrate this instability against temporal modulations within the TB using the 1D extended Su-Schrieffer-Heeger model. The Hamiltonian is given by:
\begin{eqnarray}\label{1dssh}
H =
\begin{pmatrix}
0 & 1 + \gamma_1 e^{-ik} + \gamma_2 e^{-2ik} \\
1 + \gamma_1 e^{ik} + \gamma_2 e^{2ik} & 0
\end{pmatrix}.
\end{eqnarray}
The model consists of two sublattices per unit cell, with $\gamma_1$ ($\gamma_2$) representing nearest (next-nearest) inter-cell hoppings. The model belongs to the BDI class and respects chiral symmetry, $\mathcal{C} = \sigma_z$. The phase diagram in Fig.~\ref{fig5}(a) shows different phases labeled by the integer winding number $w_1$.

For a quenched TB $(\gamma_1^L, \gamma_2^L) \rightarrow (\gamma_1^R, \gamma_2^R)$, poles in the $S$-matrix appear when the TB connects two phases with different winding numbers. Figure~\ref{fig5}(b) shows the inter-channel transmission $T_{-+}(k)$ for three different quench protocols: \textcircled{1} $(0.5, 0.5) \rightarrow (2.5, 0.5)$, \textcircled{2} $(0.5, 0.5) \rightarrow (0.5, 2)$, and \textcircled{3} $(0.5, 0.5) \rightarrow (1, 0.5)$. The RT mode is located at $k = \pi$ with perfect transfer $T_{-+}(\pi) = 1$ in case \textcircled{1}. In case \textcircled{2}, there are two such RT modes. In contrast, for case \textcircled{3}, no pole exists, and $T_{-+}(k)$ remains below unity for all momenta. For all three cases, the number of RTs align with the difference of winding numbers across the TB.  We then modulate the TB in case \textcircled{1} by assigning it a finite duration. Within the TB, two cases are studied: (i) $H^{TB}(t) = \sigma_z$, which satisfies $[H^{TB}(t), \mathcal{C}] = 0$; and (ii) $H^{TB}(t) = (1 -t / \tau_{B}) H^L + t / \tau_{B} H^R + \sigma_z$. As shown in Fig. \ref{fig5}(d), in case (i), the RT peak remains but shifts away from $\pi$; in case (ii), the RT vanishes due to the breaking of the chiral symmetry inside the TB.\\

\noindent{\large{\bf Discussion}}

\noindent To conclude, we have developed a unified framework for temporal scattering of Bloch waves at generic TBs. By drawing a close analogy between spatial and temporal scattering, we identify topologically protected RTs that emerge when the time-varying Hamiltonian undergoes a topological transition across the TB. These RTs, corresponding to poles of the $S$-matrix, give perfect inter-channel state transfer, vanishing birefringence, and a dynamically frozen quantum state. We establish a bulk-time-boundary correspondence, showing that the number of RTs is fixed by the jump of topological invariants across the TB, and in 1D,  satisfies a time-domain Levinson’s theorem. We also reveal a striking even–odd effect in the topological stability of RTs in different spatial dimensions. Our work offers a fundamentally new perspective on time-varying physics by treating the temporal region as a genuine scattering interface between two static bulk phases. This shift enables the study of nonequilibrium quantum dynamics within a standard scattering framework.

Our theory bridges temporal and spatial scattering and opens new avenues for steering quantum systems through temporal engineering. It reveals how well-designed TBs can selectively enhance, suppress, or freeze specific modes, enabling precise control of quantum dynamics. While we focus on integer AZ classes, it is promising to extend the framework to more intricate topological systems, e.g., $\mathbb{Z}_2$ AZ classes \cite{class1,class2,class3}, topological crystalline insulators/superconductors \cite{class4,tci_fu,tci_rev}, and higher-order topological phases \cite{Benalcazar2017_Science,Schindler2018_SciAdv,Langbehn2017_PRL}. Furthermore, the scattering dynamics and RTs serve as a tantalizing tool for identifying topological phases. For instance, in cold atoms, a TB can be placed between a known trivial system (e.g., with large detunings or deep lattices) and the target system. According to the bulk–time–boundary correspondence, RTs appear when the target system is topological. They manifest as the vanishing birefringence in the scattering pattern; and dynamical freezing can be detected via time-resolved Bloch-state tomography \cite{tomo1,tomo2,tomo3,tomo4}. Since RTs typically occur at high-symmetry points, a few momentum-resolved measurements suffice to determine the phase. This is broadly applicable and arguably one of the most convenient probe of topology, compared with: (i) static approaches requiring well-prepared ground states and full BZ scans; (ii) topological response relying on transport setups tailored to specific phases such as Chern insulators; or (iii) dynamical schemes such as polarization measurements \cite{qtopo7} on the $(d-1)$D band-inversion surfaces \cite{qtopo8a}.\\

\noindent{\large{\bf Methods}}

\noindent{\bf Resonant transmission in generic multiband systems.}~In the main text, we present a heuristic topological argument for the bulk-time-boundary correspondence, which states that the number of resonant transmissions (RTs) is given by the jump in topological invariants across the time boundary (TB), i.e., Eq. (\ref{rtnumber}). This argument relies on an elementary Clifford representation of the Bloch Hamiltonians. Here, we show that this identity holds in generic multiband systems, independent of any Clifford representation. Let $n_L$ and $n_R$ denote the topological invariants on either side of the TB, i.e., winding numbers in odd spatial dimensions and Chern numbers in even spatial dimensions. The basic setup is a generic $2N$-band system with a quenched TB. For both $H^L$ and $H^R$, a band gap separates the valence bands (labeled from $1$ to $N$) from the conduction bands (labeled from $N+1$ to $2N$). We set the Fermi level to zero; otherwise one can shift the overall energy without changing the eigenstates, and hence without affecting the transfer matrix $M$ or the scattering matrix $S$. 

Let us expand the Hamiltonian in the eigenbasis as $H^{L,R} = \sum_{j=1}^{2N} E^{L,R}_j|\psi^{L,R}_j\rangle \langle \psi^{L,R}_j |$. The spectra can be ``flattened” without closing the gap or altering the eigenstates $\tilde{H}^{L,R} = \sum_{j=1}^N (-1)  |\psi^{L,R}_j\rangle \langle \psi^{L,R}_j |+ \sum_{j=N+1}^{2N} (+1) |\psi^{L,R}_j\rangle \langle \psi^{L,R}_j|$. In terms of the projection operators onto the conduction and valence subspaces: $P^{L,R}_+ = \sum_{j=N+1}^{2N} |\psi^{L,R}_j\rangle \langle \psi^{L,R}_j |$, $P^{L,R}_- = 1 - P^{L,R}_+$, the flattened Hamiltonian is
\begin{equation}\label{proj1}
\tilde{H}^{L,R} = P^{L,R}_+ - P^{L,R}_- = 2P^{L,R}_+ - 1.
\end{equation}
The above procedure leaves $M$ and $S$ unchanged and preserves the topology. Consider an interpolation between the two flattened Hamiltonians, $\tilde{H}_\lambda = (1-\lambda) \tilde{H}^L+ \lambda \tilde{H}^R$, $\lambda \in [0,1]$. As $\lambda$ varies from $0$ to $1$, the topological invariant changes from $n_L$ to $n_R$ (by assumption $n_L \neq n_R$). As the topology can only change when the energy gap between the conduction and valence bands closes, there must be $|n_L - n_R|$ nontrivial gap closings along the interpolation. If a band closing involves higher-order degeneracies (e.g., quadratic), its multiplicity should be counted. 

Suppose a gap closing occurs at $\lambda=\lambda_0 \in (0,1)$ and $\bm{k}=\bm{k}_0$. At this point, there exists a nonzero state $\phi$ such that
\begin{equation} \label{gapclosing}
\tilde{H}_{\lambda_0}(\bm{k}_0) \, \phi = 0.
\end{equation}
The choice of $\phi$ is not unique, and the degree of freedom comes from the superposition in the degenerate subspace. Substituting Eq. (\ref{proj1}) into Eq. (\ref{gapclosing}), we have
\begin{equation}\label{keyproj}
 (1-\lambda_0) P^L_+ \phi+\lambda_0 P^R_+ \phi  = \frac12 \phi.
\end{equation}
Let $x = P^L_+ \phi$ and $y = P^R_+ \phi$, and applying $P^R_+$ to the left side, Eq.~\eqref{keyproj} becomes
\begin{equation}
 (1-\lambda_0) P^R_+ x+(\lambda_0-\frac{1}{2})y  = 0.
\label{eq:leftproj}
\end{equation}
Next, we show that the gap closing necessarily implies \(\det M_{++}=0\), i.e., a pole of the $S$-matrix. Here $M_{++}$ is the conduction–conduction block of the transfer matrix. Let us expand $x$ in the basis of left conduction states, \(x = \sum_{j=1}^N c_j\, |\psi^L_{N+j}\rangle\). Then in the basis of the right conduction states, we have $\langle \psi^R_{N+i} | P^R_+ x \rangle= \sum_{j=1}^N c_j\, \langle \psi^R_{N+i} | \psi^L_{N+j} \rangle= \sum_{j=1}^N (M_{++})_{ij} c_j$. 

We discuss two cases of $\lambda_0$ in Eq. (\ref{eq:leftproj}). If $\lambda_0=1/2$, we have $P^R_+x=0$, which immediately implies that $\det M_{++}=0$. If $\lambda_0\neq 1/2$, then $y=\frac{\lambda_0-1}{\lambda_0 - \frac12} P^R_+ x $. In this case, we show that  $M_{++}$ is non-invertible by contradiction. If \(M_{++}\) were invertible, \(x\) and \(y\) would be linearly independent. Applying $P^L_+$ to the left side of Eq. (\ref{gapclosing}) yields $(\tfrac12 - \lambda_0)\, x + \lambda_0\, P^L_+ y = 0$. Note that \(P^R_+ P^L_+ = M_{++}^\dagger M_{++}\) on the left conduction bands, we have $x=\frac{\lambda_0(\lambda_0-1)}{(\lambda_0-1/2)^2}P^R_+P^L_+x=\frac{\lambda_0(\lambda_0-1)}{(\lambda_0-1/2)^2}(M^{\dagger}_{++}M_{++})x$. Since \(M_{++}^\dagger M_{++}\) is positive definite and $0<\lambda_0<1$, the only possibility is the trivial solution  \(x=0\), which forces \(y=0\) and $\phi=0$ according to Eq. (\ref{keyproj}).  Therefore, for a nontrivial gap closing, \(M_{++}\) must be singular and $\det M_{++}=0$. Near $\bm{k}_0$, the block $M_{++}(\bm{k})$ is a smooth matrix-valued function of $\bm{k}$ whose rank drops at $\bm{k}_0$. If the band closing has order $m$ in the sense that the low-energy Hamiltonian first develops nonzero terms at order $(\delta k)^m$, the smallest nonzero singular value of $M_{++}(\bm{k})$ scales as $\sim (\delta k)^m$. This is nothing but the pole order $p_{\bm k_0}$ of the scattering matrix at $\bm k_0$. Summing up, we have proved the identity Eq. (\ref{rtnumber}). \\

\noindent{\bf Proof of time-domain Levinson's theorem.}~The time-domain Levinson’s theorem [Eq. (\ref{winding1})] relates the scattering phase shifts to the number of RT.  The phase shifts are given by the diagonal elements of the $S$-matrix in Eq. (\ref{Smatrix}). For block matrices, we have $\det M=\det M_{++}\det(M_{--}-M_{-+}M^{-1}_{++}M_{+-})$. The relative phase shift between the valence and conduction bands is therefore $\delta_{rel}(k)=\arg\det(M_{--}-M_{-+}M^{-1}_{++}M_{+-})-\arg\det(M_{++}^{-1})=\arg\det M(k)$. Let $U_L(k)$ and $U_R(k)$ be the unitary matrices whose columns are the eigenstates of $H^L(k)$ and $H^R(k)$, ordered from valence to conduction bands. The transfer matrix is then $M(k) = U_R^\dagger(k)U_L(k)$. It is unitary, $M^\dagger M = M M^{\dagger}=I$, $\det M(k) = \det U_R^\dagger(k) \det U_L(k)= e^{-i\phi^R(k)} e^{i\phi^L(k)}$, where $\phi^{L,R}(k) = \arg \det U_{L,R}(k)$. The relative phase shift is $\delta_{\mathrm{rel}}(k) = \phi_L(k) - \phi_R(k)$. Integrating over $k$, $\frac{1}{2\pi} \int_{-\pi}^{\pi} \partial_k \delta_{\mathrm{rel}}(k) \, dk
= \frac{1}{2\pi} \int_{-\pi}^{\pi} \partial_k \phi_L(k) \, dk
 - \frac{1}{2\pi} \int_{-\pi}^{\pi} \partial_k \phi_R(k) \, dk.$
In a smooth gauge, each term on the right side corresponds to the 1D winding number of the respective phase at the two sides of the TB: $n_{L} = \frac{1}{2\pi} \int_{-\pi}^{\pi} \partial_k \phi^L(k) \, dk$; $n_{R} = \frac{1}{2\pi} \int_{-\pi}^{\pi} \partial_k \phi^R(k) \, dk$. Thus, the integrated relative phase shift equals the difference of the winding numbers of the two phases. Combining this with Eq. (\ref{rtnumber}), we arrive at the time-domain Levinson's theorem.\\

\noindent{\large{\bf Data availability}}

\noindent  All data is available upon resonable request to the corresponding author.\\

\noindent{\large{\bf Acknowledgements}}

\noindent This work is supported by the National Key Research and Development Program of China (Grants No. 2022YFA1405800 and No. 2023YFA1406704) and National Natural Science Foundation of China (Grant No. 12474496 and No. 12547107).\\

\noindent{\large{\bf Author contributions}}

\noindent The author conceived the project, developed the theoretical framework, performed all analytical and numerical calculations, prepared the figures, and wrote the manuscript.\\

\noindent {\large{\bf Competing interests}} 

\noindent The author declares no competing interests.

\clearpage
\appendix
\setcounter{equation}{0}  %
\setcounter{figure}{0}
\counterwithout{equation}{section}  %
\renewcommand{\thefigure}{S\arabic{figure}}
\renewcommand{\thesection}{S\arabic{section}}
\pagebreak
\renewcommand{\theequation}{S\arabic{equation}}
\widetext
\begin{center}
\textbf{\large  Supplementary Material}
\end{center}

\begin{center}
\vspace{0.5em} 
Haiping Hu\textsuperscript{1, 2}\email{hhu@iphy.ac.cn}

\vspace{0.3em}
\textsuperscript{1}Beijing National Laboratory for Condensed Matter Physics, Institute of Physics, Chinese Academy of Sciences, Beijing 100190, China\\
\textsuperscript{2}School of Physical Sciences, University of Chinese Academy of Sciences, Beijing 100049, China
\end{center}
\vspace{3\baselineskip}

Appendix S1. Examples of resonant transmission (RT) as a higher-order pole;

Appendix S2. RT in the 3D chiral topological insulator.

\section{Examples of RT as a higher-order pole}
In this appendix, we illustrate the existence of higher-order resonant transmissions (RTs) in an extended 2D Qi-Wu-Zhang model. The Hamiltonian is $H(\bm k)=\bm h(\bm k)\cdot\bm\sigma$, with $h_x+ih_y=(\sin k_x+i\sin k_y)^p$ and $h_z=m-\cos k_x-\cos k_y$. The $p=1$ case reduces to the standard Qi-Wu-Zhang model in the main text. The Chern number of the lower band is $C_2=p$ when $0<m<2$; $C_2=-p$ when $-2 < m < 0$; and $C_2 = 0$ otherwise. Now we introduce a quenched TB between two Hamiltonians $H^L$ ($0<m_L<2$) with Chern number $C_2=p$ and $H^R$ ($m_R>2$) with Chern number $C_2=0$. In what follows, we take $p=2$ and locate the poles of the scattering matrix by solving $M_{++}=0$. The eigenstate of the upper band is
\begin{eqnarray}
|\psi_+(\mathbf{k})\rangle = \begin{pmatrix}
\cos \frac{\theta}{2} \\
\sin \frac{\theta}{2} e^{i\phi}
\end{pmatrix},
\end{eqnarray}
where  $\cos \theta = \frac{h_z}{|\bm h|}$, $\tan \phi = \frac{h_y}{h_x}$, and $\bm h=(h_x,h_y,h_z)=(\sin^2 k_x - \sin^2 k_y,2 \sin k_x \sin k_y, m - \cos k_x - \cos k_y)$. Thus $M_{++}=\cos\frac{\theta^L-\theta^R}{2}$. It follows that there is a single pole at $\bm k=(0,0)$. For small $\bm k$ near the pole, we have $h_x = k_x^2 - k_y^2+o(k^2)$, $h_y = 2 k_x k_y+o(k^2)$ and $h_z = m - 2 + \frac{k_x^2 + k_y^2}{2}+o(k^2)$. The angles are $\theta^L=\pi-\frac{k_x^2+k_y^2}{|m_L-2|}$ and $\theta^R=\frac{k_x^2+k_y^2}{|m_R-2|}$. Therefore,
\begin{eqnarray}
M_{++}=\frac{k_x^2+k_y^2}{2}(\frac{1}{|m_L-2|}+\frac{1}{|m_R-2|})+o(k^2).
\end{eqnarray} This is a second-order pole. For generic $p$, one can identify a $p$-th order pole at $\bm k=(0,0)$.

\section{RT in the 3D chiral topological insulator}
In this appendix, we identify the appearance of RT in a four-band systems. We consider the 3D chiral topological insulator described by the minimal Hamiltonian \cite{lxj_chiral3D}:
\begin{equation}
H_{\mathrm{3D}}(\mathbf{k}) = \left( m - \cos k_x - \cos k_y - \cos k_z \right) \gamma_0 
+ \sin k_x \, \gamma_1 + \sin k_y \, \gamma_2 + \sin k_z \, \gamma_3,
\label{eq:H3D_corrected}
\end{equation}
where the lattice momentum is $\mathbf{k} = (k_x, k_y, k_z)$, and the $\gamma$ matrices are given by
\begin{equation}
\gamma_0 = \sigma_z \otimes \tau_z, \quad
\gamma_1 = \sigma_x \otimes I, \quad
\gamma_2 = \sigma_y \otimes I, \quad
\gamma_3 = \sigma_z \otimes \tau_x.
\end{equation}
$\sigma_i$ and $\tau_i$ ($i=x,y,z$) denote two independent sets of Pauli matrices. These $\gamma$ matrices satisfy $\{\gamma_i,\gamma_j\}=2\delta_{ij} I_4~(i,j=0,1,2,3)$. The Hamiltonian has a chiral symmetry $\mathcal{C}=\sigma_z \otimes \tau_y$ satisfying $\{H_{\mathrm{3D}}(\mathbf{k}), \mathcal{C} \} = 0$. This places the model in the AIII class of the Altland–Zirnbauer (AZ) tenfold way. Its topological invariant is the 3D winding number $w_3$, giving by
\begin{equation}
w_3 =
\begin{cases}
1, & \quad 1 < m < 3, \\
-2, & \quad -1 < m < 1, \\
1, & \quad -3 < m < -1, \\
0, & \quad \text{otherwise}.
\end{cases}
\end{equation}
The bulk gap closes at the phase transition points $m = \pm 1, \pm 3$.

Let us now consider a quenched TB between $1<m_L<3$ and $m_R>3$. It is straightforward to check that a pole appears at $\bm k=(0,0,0)$. Expanding the Hamiltonian near $\mathbf{k}=\mathbf{0}$ (to linear order), we have:
\begin{equation}
H_{\mathrm{3D}}(\mathbf{k}) =
\begin{pmatrix}
\delta\,\tau_z + k_z \tau_x & (k_x - i k_y) I_2 \\
(k_x + i k_y) I_2 & -\delta\,\tau_z - k_z \tau_x
\end{pmatrix},
\label{eq:H_block}
\end{equation}
where $\delta=m-3$. The spectrum is $E_\pm(\mathbf{k}) = \pm \sqrt{\delta^2 + k^2}$. For $\delta > 0$, the four normalized eigenstates expanded in $\bm k$ are ($\pm$ labels the conduction/valence band)
\begin{align}
|\psi_{+,1}\rangle\approx
\begin{pmatrix}
1 \\
\frac{k_z}{2\delta} \\
\frac{k_x + i k_y}{2\delta} \\
0
\end{pmatrix},~~~
|\psi_{+,2}\rangle\approx
\begin{pmatrix}
0 \\
\frac{k_x - i k_y}{2\delta} \\
-\frac{k_z}{2\delta} \\
1
\end{pmatrix};~~~
|\psi_{-,1}\rangle\approx
\begin{pmatrix}
-\frac{k_z}{2\delta} \\
1 \\
0 \\
-\frac{k_x + i k_y}{2\delta}
\end{pmatrix},~~~
|\psi_{-,2}\rangle \approx
\begin{pmatrix}
-\frac{k_x - i k_y}{2\delta} \\
0 \\
1 \\
\frac{k_z}{2\delta}
\end{pmatrix}.
\label{eq:neg_states}
\end{align}
If $\delta < 0$, the conduction and valence bands are exchanged. For our TB setting, we have $\delta_L=m_L-3<0$ and $\delta_R=m_R-3>0$.  The conduction–conduction overlap matrix $M_{++}(\mathbf{k})$ is given by
\begin{equation}
\big[M_{++}(\mathbf{k})\big]_{mn} =
\langle \psi^R_{+,m} | \psi^L_{+,n} \rangle, \quad m,n=1,2.
\end{equation}
To leading order in $\bm k$, we find
\begin{equation}
M_{++}(\mathbf{k}) =(\frac{1}{2\delta_R}-\frac{1}{2\delta_L})
\begin{pmatrix}
k_z & k_x - i k_y \\
k_x + i k_y & -k_z
\end{pmatrix}
+ \mathcal{O}(k^2).
\label{eq:Mpp_result}
\end{equation}
It is clear that the $M_{++}(\mathbf{k})$ block takes the Weyl form $M_{++}(\bm k)\sim\bm k\cdot\bm\sigma$. This is a first-order pole of the scattering S-matrix, consistent with the difference of topological invariant ($w_{3,L}$=1 and $w_{3,R}=0$).

\end{document}